\documentclass[seceq]{ptptex}

\usepackage{graphicx}
\usepackage{bm}

\newcommand{\PDG}{Nakamura:2010zzi}
\newcommand{\piK}{\pi^- p\to K^-\Theta^+}
\newcommand{\Kpi}{K^+ p\to \pi^+\Theta^+}
\newcommand{\Slash}[1]{\ooalign{\hfil/\hfil\crcr$#1$}}

\preprintnumber[3cm]{
J-PARC TH 008}

\title{
Meson-Induced Pentaquark Productions%
}

\author{
Tetsuo \textsc{Hyodo}$^{1}$, %
Atsushi \textsc{Hosaka}$^{2}$ and %
Makoto \textsc{Oka}$^{1,3}$%
}

\inst{
$^{1}$Department of Physics, Tokyo Institute of Technology, Tokyo 152-8551, Japan\\
$^{2}$Research Center for Nuclear Physics (RCNP),
Ibaraki 567-0047, Japan\\
$^{3}$J-PARC Branch, KEK Theory Center,
Institute of Particle and Nuclear Studies, 
High Energy Accelerator Research Organization (KEK),\\
Tokai, Ibaraki, 319-1106, Japan
}

\abst{
Production cross sections of the pentaquark $\Theta^{+}$ baryon in the meson-induced reactions are calculated for $J^{P}=1/2^{\pm}$ and $3/2^{\pm}$ cases. Through the comparison with the previous measurements at KEK, several quantum numbers are excluded. The remaining possibilities can be used to cast the upper limit on the $\Theta^{+}$ width, combining with the result from the J-PARC E19 experiment.
}

\begin{document}

\maketitle

\section{Introduction}\label{sec:intro}

Exploration of exotic hadrons which have more than three valence quarks is one of the important issues in hadron physics. Such states are not forbidden in QCD, but no exotic hadron has been established so far, in contrast to the existence of hundreds of ordinary hadrons.\cite{\PDG} It was shown in chiral dynamics that the formation of $s$-wave molecule type hadrons is less likely in exotic channels than in ordinary channels.\cite{Hyodo:2006yk,Hyodo:2006kg} This observation may partly explain the difficulty of finding the exotic hadrons in experiments. Recently, active discussion is ongoing in the heavy quark sector, stimulated by the reports of newly observed extraordinary states~\cite{Choi:2007wga,Belle:2011aa} (see also Refs.~\citen{Swanson:2006st,Brambilla:2010cs}).

One of the candidates of manifestly exotic hadrons in the light and strange quark sector is the pentaquark $\Theta^+$ (strangeness $S=+1$ baryon), reported in photoproduction experiments.\cite{Nakano:2003qx,Nakano:2008ee} Among several reaction processes to produce $\Theta^+$, hadronic reactions are in general less ambiguous than the others from the theoretical point of view. At tree level of the effective Lagrangian approach, all the coupling constants can be determined from experiments, while there arises ambiguity in the photoproduction reactions from anomalous magnetic moment of $\Theta^+$.

The $\Theta^+$ production experiments with meson beams were performed at KEK-PS. In the $\piK$ reaction,\cite{Miwa:2006if} a bump structure was observed at 1530 MeV, although the statistical significance is not enough to claim the peak as a resonance signal. In the $\Kpi$ reaction, no peak structure was found.\cite{Miwa:2007xk} In both the reactions, the upper limit of the cross section of the $\Theta^+$ formation was estimated to be of the order of several $\mu$b. Recent J-PARC E19 experiment has reported no peak structure in the $\piK$ reaction, which casts even severer upper limit of less than 1 $\mu$b.\cite{Shirotori:2012ka} These upper limits are much smaller than the typical cross section values of ordinary hadronic processes. \ For instance, the cross sections of \ the background processes in $\pi^-p \to K^+KN$ has measured to be $\sim 25$ $\mu$b (direct process), $\sim 21$ $\mu$b (via $\Lambda(1520)$ in the $\bar{K}N$ pair), and $\sim 30$ $\mu$b (via $\phi$ in the $\bar{K}K$ pair),\cite{Dahl:1967pg} and the cross section of the two-body hyperon productions are found to be 100--200 $\mu$b.\cite{Dahl:1969ap}

Here we would like to study the $\piK$ and $\Kpi$ reactions theoretically, in order to see how crucial for the existence of $\Theta^+$ the upper limits of the production cross sections are. The easiest interpretation of the non-observation of a peak in experiments is the absence of $\Theta^+$, but one has to know whether the size of the signal cross section is indeed smaller than theoretically expected. Effective Lagrangian approach to the meson-induced $\Theta^{+}$ production has been studied in Refs.~\citen{Hyodo:2003th,Liu:2003rh,Oh:2003kw,Oh:2003gj,Ko:2003xx,Hyodo:2005bw} with several reaction mechanisms and different schemes. To understand the newly obtained experimental results, it is important to revisit the reaction study for various cases of unknown quantum numbers of $\Theta^{+}$.

\begin{figure}[tbp]
    \centering
    \includegraphics[width=4cm,clip]{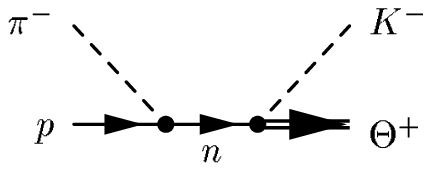}
    \includegraphics[width=4cm,clip]{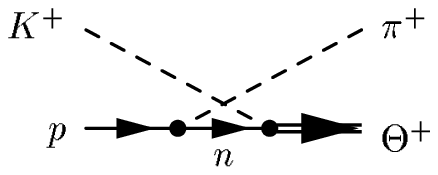}
    \caption{\label{fig:Born}
    Born diagrams for the meson-induced $\Theta^+$ production:
    $s$-channel diagram for $\piK$ (left);
    $u$-channel diagram for $\Kpi$ (right). 
    }
\end{figure}

In this paper, we consider the minimal diagrams shown in Fig.~\ref{fig:Born}: $s$-channel ($u$-channel) nucleon exchange in the $\piK$ ($\Kpi$) reaction. These diagrams include the $KN\Theta$ coupling which is responsible for $\Theta^{+}$ to decay into the $KN$ state. We assume that $\Theta^{+}$ is isosinglet, so the only $s$- or $u$-channel diagram is \\allowed for each reaction. We shall not consider the other mechanisms, such as $K^{*}$ vector meson exchange~\cite{Oh:2003kw} and unconventional two-meson couplings.\cite{Hyodo:2005bw} In general, several different diagrams could destructively interfere to produce a small cross section in a certain reaction. Such destructive interferences, however, cannot explain all the negative results of $\Theta^{+}$ in various reactions, such as photo-productions, proton-proton collisions, and meson-induced reactions. Thus, it is reasonable to assume that there is no accidental interference of several diagrams in the meson-induced reactions, and that all possible contributions are individually small. In this case, using the diagrams in Fig.~\ref{fig:Born}, we can make a conservative estimation of the upper limit of the $\Theta^{+}$ width.

\section{Theoretical framework}\label{sec:framework}

To evaluate the diagrams in Fig.~\ref{fig:Born}, we need Yukawa couplings for the $KN\Theta$ and $\pi NN$ vertices. In the effective Lagrangian approach, there are two schemes to introduce the Yukawa couplings, namely, pseudoscalar (PS) scheme and pseudovector (PV) scheme. For the $\pi NN$ coupling, the former (latter) corresponds to the linear (nonlinear) representation of chiral symmetry. We adopt both schemes to estimate theoretical uncertainties. Four possible cases for the unknown spin and parity of $\Theta^{+}$ are examined: $J^{P}=1/2^{\pm}, 3/2^{\pm}$.

For spin $1/2$ cases, the interaction Lagrangians in the PS scheme are given by~\cite{Oh:2003gj}
\begin{align}
    \mathcal{L}_{KN\Theta}^{1/2^{\pm}}
    &=g_{KN\Theta}^{1/2^{\pm}}
    \bar{\Theta}^+\Gamma KN + \text{h.c.},
    \nonumber \\
    \mathcal{L}_{\pi NN}
    &=ig_{\pi NN}\bar{N}\gamma_{5} \bm{\pi} N,
    \nonumber 
\end{align}
where $\Gamma = 1$ for negative parity and $\Gamma = i\gamma_{5}$ for positive parity. The $KN\Theta$ coupling constant is determined by the decay width $\Gamma_{\Theta}$ as
\begin{equation}
    g_{KN\Theta}^{1/2^{\pm}} = \sqrt{
    \frac{2\pi M_{\Theta}\Gamma_{\Theta}}{q(E_{N}\mp M_{N})}}  ,
    \nonumber
\end{equation}
where $M_{N}$ ($M_{\Theta}$) is the mass of the nucleon ($\Theta^{+}$), $q$ is the magnitude of the relative three-momentum of the two particles in the final state and $E_{N}=\sqrt{q^{2}+M_{N}^{2}}$. The interaction Lagrangians in the PV scheme are the derivative couplings
\begin{align}
    \mathcal{L}_{KN\Theta}^{1/2^{\pm}}
    &=\frac{-ig_{A}^{*\pm}}{2f}\bar{\Theta}^+\gamma_{\mu}\Gamma
    \partial^{\mu}KN + \text{h.c.},
    \nonumber \\*
    \mathcal{L}_{\pi NN}
    &=\frac{g_{A}}{2f}\bar{N}\gamma_{\mu}\gamma_5 
    \partial^{\mu}\bm{\pi} N ,
    \nonumber
\end{align}
where $f$ is the meson decay constant and $g_{A}$ is the axial coupling constant of the nucleon. The transition axial coupling constant for the $\Theta^+\to N$ is related to the $KN\Theta$ coupling through the generalized GT relation~\cite{Nam:2003uf} as 
\begin{align}
    g_{A}^{*\pm}
    =\frac{2f}{M_{\Theta}\pm M_{N}}g_{KN\Theta}^{1/2^{\pm}}
    \nonumber
    ,
\end{align}
which can be shown by comparing the expression of the decay width of $\Theta^{+}$ in both the schemes. The interaction Lagrangian for the spin $3/2$ cases is given by
\begin{align}
    \mathcal{L}_{KN\Theta}^{3/2^{\pm}}
    &=\frac{-ig_{KN\Theta}^{3/2^{\pm}}}{m_{K}}\bar{\Theta}^{+\mu}
    \gamma_{5}\Gamma\partial_{\mu}K N + \text{h.c.},
    \label{eq:3o2Lag}
\end{align}
where the $KN\Theta$ coupling is given by
\begin{align}
    g_{KN\Theta}^{3/2^{\pm}}
     = \sqrt{
    \frac{6\pi M_{\Theta}m_{K}^{2}\Gamma_{\Theta}}{q^{3}(E_{N}\pm M_{N})}}
    \nonumber .
\end{align}
In the spin $3/2$ case, we use the PV scheme for the $\pi NN$ coupling, in order to be consistent with the derivative coupling of Eq.~\eqref{eq:3o2Lag}.

It is straightforward to calculate the Born amplitude using these effective Lagrangians. For instance, the amplitude of the $\pi^-(k)p(p)\to K^-(k^{\prime})\Theta^+(p^{\prime})$ reaction for the $1/2^{+}$ case in the PS scheme is given by
\begin{align}
    -it_{\piK}
    &=
    \bar{u}_{\Theta^{+}}
    (-g_{KN\Theta}\gamma_5)
    \dfrac{i(\Slash{p}+\Slash{k}+M_{N})}{s-M_{N}^2}
    (-\sqrt{2}g_{\pi NN}\gamma_5 ) u_{p} .
    \nonumber 
\end{align}
In the same way, for the $K^+(k)p(p)\to \pi^+(k^{\prime})\Theta^+(p^{\prime})$ reaction, we obtain
\begin{align}
    -it_{\Kpi}
    &=
    \bar{u}_{\Theta^{+}}
    (-g_{KN\Theta}\gamma_5)
    \dfrac{i(\Slash{p}-\Slash{k}^{\prime}+M_{N})}{u-M_{N}^2}
    (-\sqrt{2}g_{\pi NN}\gamma_5 )u_{p} , 
    \nonumber
\end{align}
where $u=(p-k^{\prime})^2$.

Using the kinematics in the center-of-mass frame, the amplitude square can be expressed as a function of the total energy $\sqrt{s}$ and scattering angle $\theta$. The differential cross section can be written as
\begin{align}
    \frac{d\sigma}{d\Omega} (\sqrt{s},\cos\theta) 
    =&\frac{1}{4\pi s}\frac{|\bm{k}^{\prime}|}{|\bm{k}|}
    M_{N}M_{\Theta}
    \frac{1}{2\pi}
    \frac{1}{2}
    \cdot 
    \frac{1}{2}\sum_{\text{spin}} |t(\sqrt{s},\cos\theta)|^2 F(\sqrt{s},\cos\theta) ,
    \label{eq:dcross}
\end{align}
where we have introduced a phenomenological form factor $F(\sqrt{s},\cos\theta)$ which reflects the finite sizes of the hadrons. As in Ref.~\citen{Oh:2003kw}, we examine two types of form factors for the estimation of theoretical uncertainties. The static form factor is the three-momentum monopole type, defined as
\begin{align}
    F(\sqrt{s},\cos\theta)
    =&F_{s}^{2},\quad
    F_s
    =\frac{\Lambda_s^2}{\Lambda_s^2+|\bm{k}|^2} .
    \label{eq:FF1} 
\end{align}
The magnitude of the initial three momentum $|\bm{k}|^2$ is determined only by the total energy $\sqrt{s}$ and the masses of the initial particles, so this factor does not produce any angular dependences. The covariant form factor is also used often in photoproduction processes. The following form factor is introduced at each vertex
\begin{align}
    F(\sqrt{s},\cos\theta)
    =&F_{c}^{4},
    \quad
    F_c
    =\frac{\Lambda_c^4}{\Lambda_c^4+(x-M_{ex}^2)^2}
    \label{eq:FF2}
    ,
\end{align}
where $x$ is the Mandelstam variable of the diagram of interest and $M_{ex}$ is the mass of the exchanged particle. In the present case, $x=s$ ($x=u$) for the $\piK$ ($\Kpi$) reaction and $M_{ex} = M_{N}$ for both cases. Note that the form factor for the $\Kpi$ reaction introduces angular dependence through the variable $u$. 

\section{Numerical results}\label{sec:results}

Here we present the numerical results. We set the mass of $\Theta^{+}$ as $M_{\Theta}=1540$ MeV. Since the Born diagrams are proportional to the $KN\Theta$ coupling constant, the cross section is simply proportional to the width of $\Theta^{+}$:
\begin{align}
    \frac{d\sigma}{d\Omega}
    \propto & 
    g^{2}
    \propto \Gamma_{\Theta} .
    \nonumber
\end{align}
In the following, we present the cross sections in units of $\Gamma_{\Theta}/(1$ MeV) [$\mu$b]. The other constants are fixed at the standard values: $\pi N$ Yukawa coupling $g_{\pi NN} = 13.5$, meson decay constant $f=93$ MeV, axial coupling constant of the nucleon $g_{A} = 1.25$. The cutoff of the static form factor is chosen to be $\Lambda_s= 500$ MeV.\cite{Liu:2003rh,Oh:2003kw} The value of the cutoff in the covariant form factor is $\Lambda_c= 1800$ MeV, which was used for the calculation of the kaon photoproduction processes.\cite{Janssen:2001wk,Oh:2003kw} Later we will examine small variation of the cutoff values to examine the cutoff dependence.

\begin{figure}[tbp]
    \centering
    \includegraphics[width=9cm,clip]{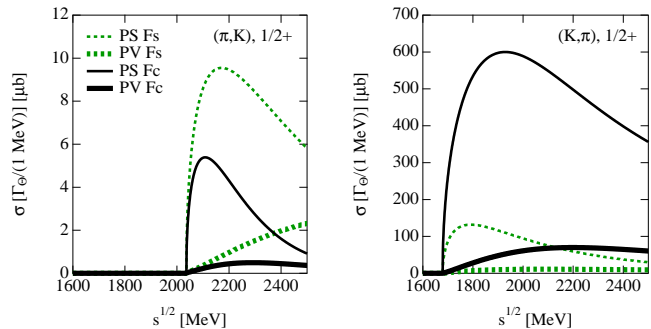}
    \caption{\label{fig:totcross_FF_pos}
    Total cross sections of the $\piK$ reaction (left) and the $\Kpi$ reaction for the $J^{P}=1/2^{+}$ case in units of $\Gamma_{\Theta}/(1$ MeV) [$\mu$b]. The form factor is chosen to be the static one with $\Lambda_s= 500$ MeV (dashed lines) and the covariant one with $\Lambda_c= 1800$ MeV (solid lines). Thin (thick) lines represent the results in the PS (PV) scheme.}
\end{figure}

In Fig.~\ref{fig:totcross_FF_pos}, we show the results of the total cross sections which are obtained by integrating the differential cross section in Eq.~\eqref{eq:dcross}. We first show the results with $J^{P}=1/2^{+}$, examining the static and covariant form factors in the PS and PV schemes. The magnitude of the cross section for the $1/2^{+}$ case is found to be several micro barn around the threshold, when the width of $\Theta^{+}$ is 1 MeV. The different threshold behaviors of the PS and PV schemes can be understood in the following way.

When $\Theta^{+}$ has $J^{P}=1/2^{+}$, then the process of interest is  a flavor analogue of the $\pi N\to \pi N$ reaction. According to the chiral low energy theorem, the meson-baryon scattering amplitude at threshold should behave as a quantity of ${\cal O}(m/M)$ or higher, where $m$ and $M$ are the masses of the NG boson and the baryon. In the PV scheme, the Born amplitude behaves as ${\cal O}(m/M)^2$ at threshold, consistent with this observation\footnote{This counting is only valid at the threshold. In general, the Born terms are counted as $\mathcal{O}(p)$ in chiral perturbation theory, and the additional $m/M$ factor appears at threshold. For the transition among baryons within the same flavor multiplet, the Weinberg-Tomozawa term contributes as ${\cal O}(m/M)$, which is absent in the $N\to \Theta^{+}$ case~\cite{Hosaka:2004mv}.}. In contrast, a naive construction of the Born amplitude in the PS scheme leads to an amplitude of ${\cal O}(1)$, which should be cancelled by the scalar meson exchange to give ${\cal O}(m/M)$ amplitude. In the $\pi N$ scattering, the scalar meson $\sigma$ is introduced as a chiral partner of $\pi$. However, in the present case, the interaction in the scalar channel is not well understood, because of the lack of information on the scalar meson in the strange sector and of the multiplicity of the chiral representation for $\Theta^+$~\cite{Beane:2004ix}. Thus, the enhancement of the PS scheme can be attributed to the ${\cal O}(1)$ amplitude in the Born diagrams. The smallness of the PV amplitude was also pointed out in Refs.~\citen{Ko:2003xx} and \citen{Hyodo:2005bw}. 

We calculate the $\Theta^{+}$ production for other quantum numbers, using effective Lagrangians in \S~\ref{sec:framework}. The results of the total cross sections are summarized in Table~\ref{tbl:totcross} where we choose the initial energy as $\sqrt{s} = 2124$ MeV (1888 MeV) for the $\piK$ ($\Kpi$) reaction. These energies correspond to those in the KEK experiments, $P_{\text{lab}}\sim 1920$ MeV and 1200 MeV, respectively. The different schemes and different form factors may be considered to be the theoretical uncertainties in the calculation. 

We have checked the dependence on the cutoff parameter of the form factor. At the above energies, 5 \% variation of $\Lambda_{s}$ in the static form factor~\eqref{eq:FF1} results in the deviation of the total cross section of about 15 \% for both the $\piK$ and $\Kpi$ reactions. For the covariant case~\eqref{eq:FF2}, 5 \% variation of $\Lambda_{c}$ leads to\\ 45 \% (5 \%) deviation of the total cross section for the $\piK$ ($\Kpi$) reaction. The dependence of the cross section on the cutoff values is also shown\\ in Table~\ref{tbl:totcross}. Since the covariant form factor is introduced at each vertex, the cross section is more sensitive to the cutoff value than the static one. This is, however, not the case for the $\Kpi$ reaction, because of the small off-shellness of the intermediate nucleon at this kinematics.

At this point, we check the dominance of the Born diagrams by calculating the cross section for the hyperon production $\pi^{-} p \to K^{+} \Sigma^{-}$ with similar initial energy. Since this is a double-charge exchange reaction, there is no $t$-channel exchange contribution as well as the Weinberg-Tomozawa contact interaction in the PV scheme. We apply the present model with $s$- and $u$-channel Born diagrams to this reaction using the hyperon couplings determined by the $SU(3)$ symmetry. At $P_{\text{lab}}\sim $ 1940 MeV, we obtain $\sim 200$ $\mu$b ($1200$--$1800$ $\mu$b) in the PS (PV) scheme. Although the experimental value of $98\pm 10$ $\mu$b\cite{Dahl:1969ap} is not perfectly reproduced (the discrepancy can be reduced by modifying the form factor), it is clear that the nucleon Born terms give a large contribution to the cross section. We thus consider that the present model contains the important contributions of this reaction.

From Table~\ref{tbl:totcross}, we observe $\sigma_{1/2^{-}}< \sigma_{1/2^{+}},\sigma_{3/2^{+}}< \sigma_{3/2^{-}}$ as a general tendency. This can be understood by the relation between the $KN\Theta$ coupling constant $g$ and the partial wave $l$ of the $KN$ decay. For $\Gamma_{\Theta} \propto g^{2}(q/M_{\Theta})^{2l+1}$ ($q\sim 268$ MeV)\cite{Hyodo:2005wa} being fixed, the higher $l$ corresponds to the larger coupling constant. Thus, the larger cross section is obtained when the higher $l$ is assigned to the $\Theta^{+}$ decay. Some exceptional behaviors have specific reasons. For instance, large cross sections in the PS scheme for the $1/2^{+}$ can be understood by the lack of the scalar meson exchange as explained above. In the $3/2^{-}$ case, there is a large difference between $\piK$ and $\Kpi$ reactions, which is related to the partial wave of the final states. In the $\piK$ reaction, since the nucleon is exchanged in $s$-channel, the total $J^{P}$ of the system is restricted to be $1/2^{+}$. As a consequence, the final state of $K^{-}$ ($0^{-}$) and $\Theta^{+}$ ($3/2^{-}$) should be in $d$ wave, which suppresses the cross section. On the other hand, the $\Kpi$ reaction has no such restriction to the total $J^{P}$, so the final $K^{-}\Theta^{+}$ pair can be in $s$ wave (total $J^{P}=3/2^{+}$), provided that the initial state is in $l=1$. This explains the difference of the cross section in the $3/2^{-}$ case.

\begin{table}[tbp]
    \centering
    \caption{Total cross sections of the $\piK$ reaction at $\sqrt{s} = 2124$ MeV ($P_{\text{lab}}\sim 1920$ MeV)
    and the $\Kpi$ reaction at $\sqrt{s} = 1888$ MeV ($P_{\text{lab}}\sim 1200$ MeV). Cutoff values are chosen to be $\Lambda_s= 500$ MeV ($\Lambda_c= 1800$ MeV) for the static (covariant) form factors, with uncertainties that arise from 5\% variation of the cutoff values. All the numbers are given in units of $\Gamma_{\Theta}/(1$ MeV) [$\mu$b].}
    \begin{tabular}{|c|ll|ll|}
        \hline
         &\multicolumn{2}{l|}{$\piK$}  & \multicolumn{2}{l|}{$\Kpi$}   \\ \hline
        $J^{P}=1/2^{+}$ & PS  & PV  & PS & PV  \\
        \hline
        static 
    & \phantom{0}9.2\phantom{0}$^{+1.4}_{-1.3}$ & 0.51$^{+0.07}_{-0.08}$ 
    & \phantom{00}119\phantom{.0}$^{+14}_{-14}$ & \phantom{0}9.6$^{+1.1}_{-1.1}$  \\
        covariant
	& \phantom{0}5.3\phantom{0}$^{+2.8}_{-2.0}$ & 0.29$^{+0.16}_{-0.11}$ 
	& \phantom{00}595\phantom{.0}$^{+16}_{-20}$  & 46\phantom{.0}$^{+1}_{-2}$     \\
        \hline
        $J^{P}=1/2^{-}$ & PS  & PV  & PS  & PV   \\
        \hline
        static 
    & \phantom{0}0.18$^{+0.02}_{-0.03}$ & 0.40$^{+0.06}_{-0.06}$ 
    & \phantom{0000}1.9$^{+0.3}_{-0.2}$ & \phantom{0}4.2$^{+0.5}_{-0.5}$  \\
        covariant
	& \phantom{0}0.10$^{+0.06}_{-0.04}$ & 0.23$^{+0.12}_{-0.09}$ 
	&  \phantom{0000}9.6$^{+0.3}_{-0.3}$ & 20\phantom{.0}$^{+1}_{-1}$    \\
        \hline
        $J^{P}=3/2^{+}$ & & & &  \\
        \hline
        static
    & 10\phantom{.00}$^{+2}_{-1}$ & 
    & \phantom{000}94\phantom{.0}$^{+11}_{-11}$ &   \\
        covariant
	&  \phantom{0}5.9\phantom{0}$^{+3.1}_{-2.2}$ &  
    &  \phantom{00}478\phantom{.0}$^{+12}_{-14}$ &     \\
        \hline
        $J^{P}=3/2^{-}$ & & & &  \\
        \hline
        static
    & \phantom{0}5.5\phantom{0}$^{+0.8}_{-0.8}$ & 
    & \phantom{0}8572\phantom{.0}$^{+1019}_{-992}$ &   \\
        covariant
	&  \phantom{0}3.2\phantom{0}$^{+1.6}_{-1.2}$ & 
    &  40544\phantom{.0}$^{+1511}_{-1824}$ &     \\
        \hline
    \end{tabular}
    \label{tbl:totcross}
\end{table}

\begin{figure}[tbp]
    \centering
    \includegraphics[width=9cm,clip]{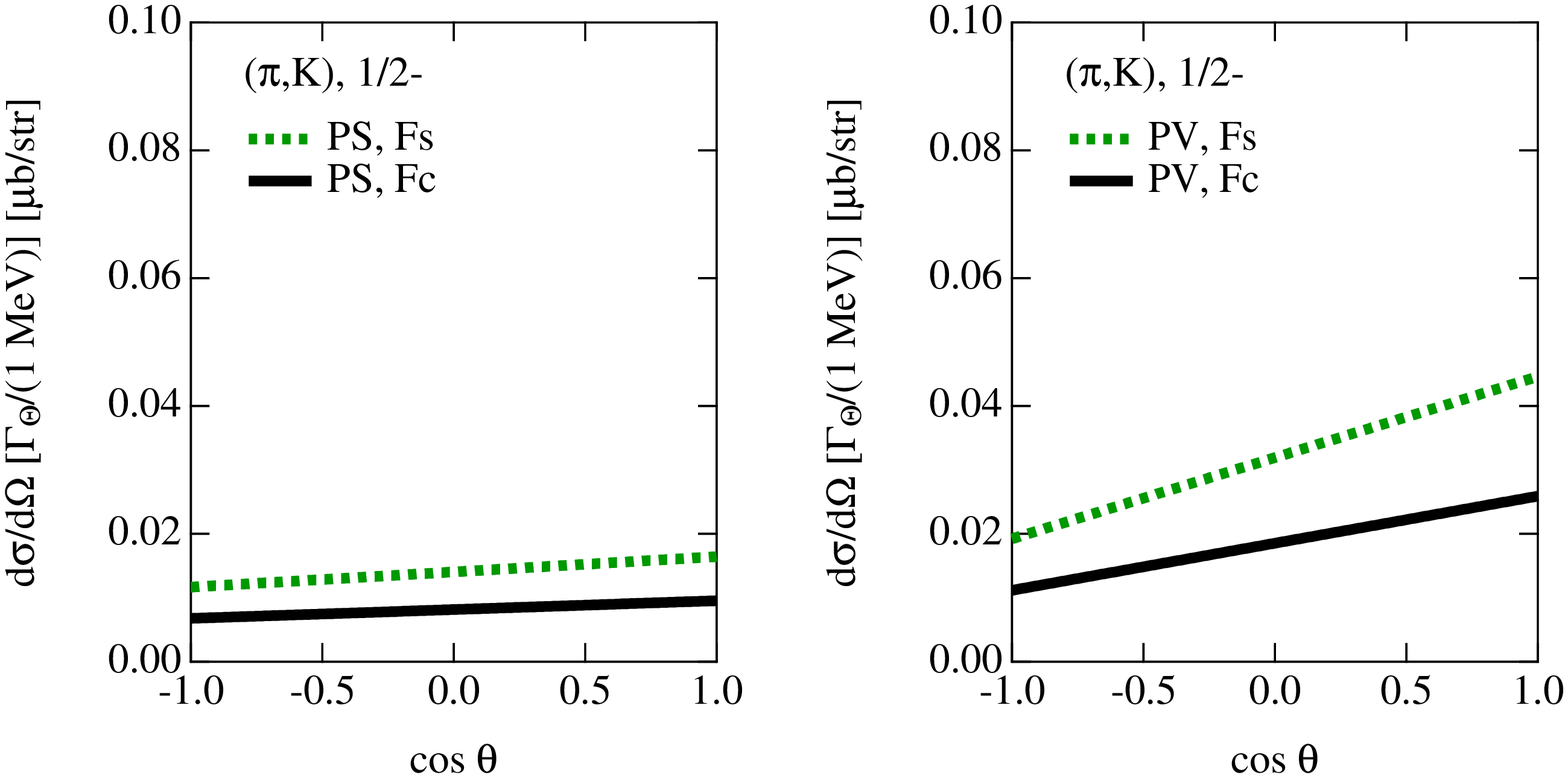}
    \includegraphics[width=4.5cm,clip]{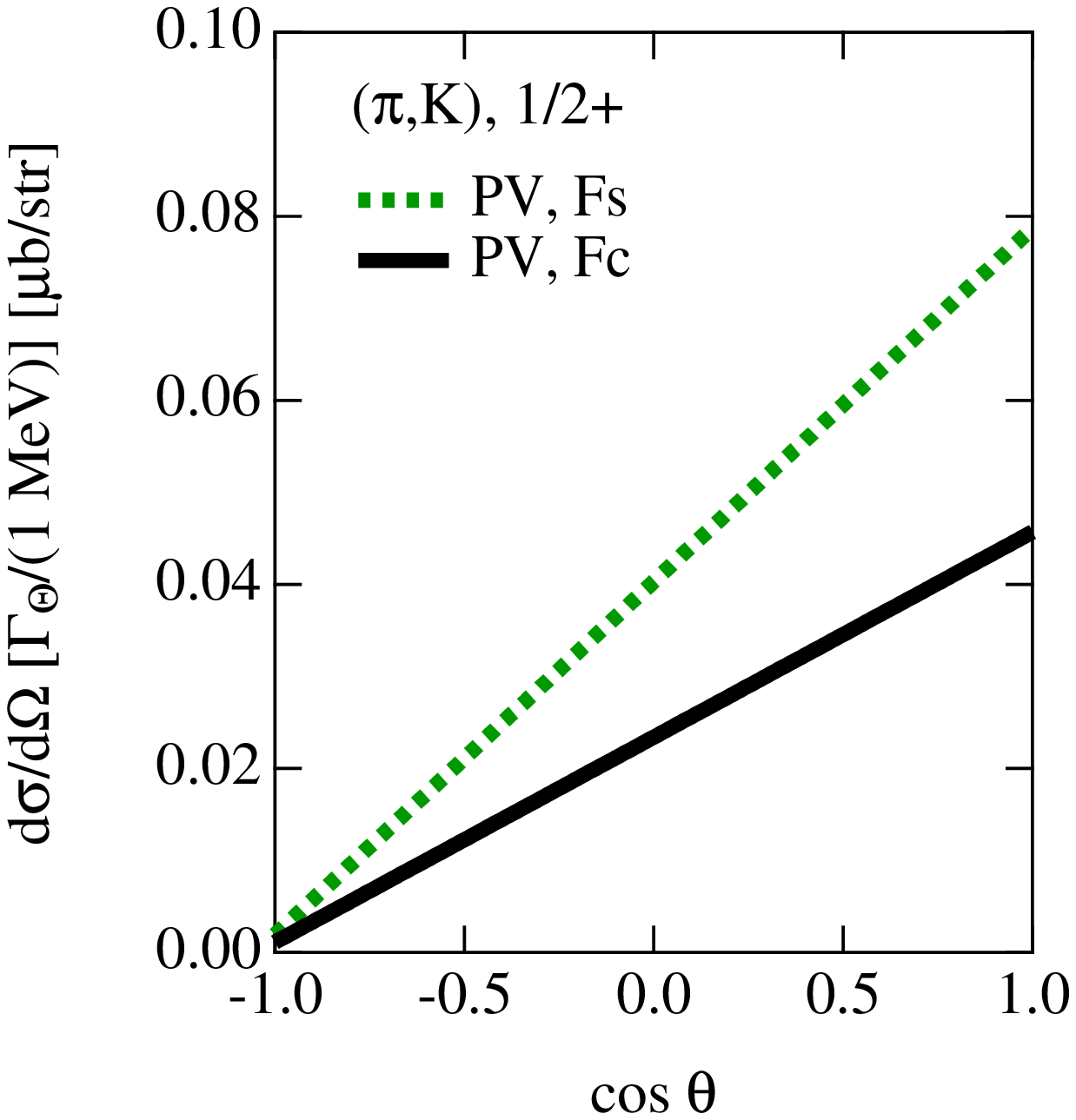}
    \caption{\label{fig:diffcross}
    Differential cross sections of the $\piK$ reaction with the PS scheme for $1/2^{-}$ case (left), the PV scheme for $1/2^{-}$ case (middle), and the PV scheme for $1/2^{+}$ case (right) in units of $\Gamma_{\Theta}/(1$ MeV) [$\mu$b/str]. The form factor is chosen to be the static one with $\Lambda_s= 500$ MeV (dashed lines) and the covariant one with $\Lambda_c= 1800$ MeV (solid lines). }
\end{figure}%

The KEK experiments gave the upper limits of the cross sections of the order of several $\mu $b,\cite{Miwa:2006if,Miwa:2007xk} and J-PARC E19 gives the upper limit of less than 1 $\mu$b for the $\piK$ cross section.\cite{Shirotori:2012ka} In order to be consistent with the cross sections of the $\Kpi$ reaction with spin $3/2$ cases in Table~\ref{tbl:totcross}, $\Gamma_{\Theta}$ must be much smaller than 1 MeV. Considering that the typical width of strongly decaying particles is larger than 1 MeV, the spin 3/2 cases are almost excluded by the results in Table~\ref{tbl:totcross}. Small cross sections of the $\piK$ reaction can be achieved by the PV scheme for $1/2^{+}$ case and both schemes for $1/2^{-}$ case.\footnote{Note, however, that the $\Theta^{+}$ decays into $KN$ channel in $s$-wave in the $1/2^{-}$ case. From the viewpoint of hadron structure, the width less than 1 MeV in $s$-wave decay (with appreciable phase space) is highly unlikely.\cite{Hosaka:2004bn}} In Fig.~\ref{fig:diffcross}, we show the angular dependence of the differential cross sections for these cases. The angular dependence is weak for the $1/2^{-}$ cases, while some forward peak structure is observed for the $1/2^{+}$ case. This information will be useful for the comparison with the experimental results with limited angular coverage. Using the new result by J-PARC E19 experiment and the central values of the theoretical prediction, the upper limit of the $\Theta^{+}$ width is obtained as 0.72 MeV (3.1 MeV) for $1/2^{+}$ ($1/2^{-}$) $\Theta^{+}$.\cite{Shirotori:2012ka}

\section{Summary}\label{sec:summary}

We have estimated the cross sections of the $\piK$ and $\Kpi$ reactions using the effective Lagrangian approach. The $\Theta^{+}$ quantum numbers up to spin 3/2 are examined. To explain the small cross sections in both $\piK$ and $\Kpi$ reactions reported by KEK experiments~\cite{Miwa:2006if,Miwa:2007xk}, the spin $3/2$ possibility is highly disfavored. We present theoretical estimates of the cross section and its angular dependence for the spin 1/2 cases, under the assumption that the nucleon Born diagrams are dominant and the effect of the destructive interference with other contributions is small. Combining the central value of the theoretical prediction 
with the severer constraint by the J-PARC E19 experiment, the upper limit of the $\Theta^{+}$ width is obtained as 0.72 MeV (3.1 MeV) for the $J^{P}=1/2^{+}$ ($1/2^{-}$) case.\cite{Shirotori:2012ka}

\section*{Acknowledgements}
We would like to thank Megumi Naruki, Kotaro Shirotori, Tomonori Takahashi, and Toshiyuki Takahashi for discussion. 
T.H. thanks the support from the Global Center of Excellence Program by MEXT, Japan, through the Nanoscience and Quantum Physics Project of the Tokyo Institute 
of Technology. 
A.H. is supported in part by the Grant-in-Aid for Scientific
Research on Priority Areas entitled ``Elucidation of New
Hadrons with a Variety of Flavors'' (E01: 21105006).
This work is partly supported by the Grant-in-Aid for Scientific Research from 
MEXT and JSPS (Nos.\
  19540275, 
  21840026, 
  and 22105503).
%

%




\end{document}